\documentclass[showpacs,preprintnumbers,superscriptaddress,prl,twocolumn]{revtex4-1}

\usepackage{graphicx}
\usepackage{amsmath}
\usepackage{times}

\newcommand{\op}[1]{\hat{#1}}
\newcommand{\cre}[1]{\hat{#1}^\dagger}
\newcommand{\des}[1]{\hat{#1}}

\newcommand{\Pu}{\mathrm{p}}
\newcommand{\Sp}{\mathrm{s}}
\newcommand{\M}{\mathrm{m}}

\newcommand{\C}{\mathrm{r}}

\newcommand{\ccr}{\hat{c}^\dagger}
\newcommand{\cde}{\hat{c}}
\newcommand{\acr}{\hat{a}^\dagger}
\newcommand{\ade}{\hat{a}}

\newcommand{\xihat}{\hat{\xi}}

\newcommand{\im}{i}
\newcommand{\ex}{e}

\newcommand{\therm}{\mathrm{th}}

\begin{document}

\title[Signatures of nonlinear cavity optomechanics in the weak coupling regime]{Signatures of nonlinear cavity optomechanics in the weak coupling regime}
\author{K. B{\o}rkje}
\affiliation{Niels Bohr Institute, University of Copenhagen, Blegdamsvej 17, DK-2100 Copenhagen, Denmark}
\author{A. Nunnenkamp}
\affiliation{Department of Physics, University of Basel, Klingelbergstrasse 82, CH-4056 Basel, Switzerland}
\author{J. D. Teufel}
\affiliation{National Institute of Standards and Technology, Boulder, Colorado 80305, USA}
\author{S. M. Girvin}
\affiliation{Departments of Physics and Applied Physics, Yale University, New Haven, Connecticut 06520, USA}

\date{\today}

\begin{abstract}
We identify signatures of the intrinsic nonlinear interaction between light and mechanical motion in cavity optomechanical systems. These signatures are observable even when the cavity linewidth exceeds the optomechanical coupling rate. A strong laser drive red-detuned by twice the mechanical frequency from the cavity resonance frequency makes two-phonon processes resonant, which leads to a nonlinear version of optomechanically induced transparency. This effect provides a new method of measuring the average phonon number of the mechanical oscillator. Furthermore, we show that if the strong laser drive is detuned by half the mechanical frequency, optomechanically induced transparency also occurs due to resonant two-photon processes. The cavity response to a second probe drive is in this case nonlinear in the probe power. These effects should be observable with optomechanical coupling strengths that have already been realized in experiments.
\end{abstract}


\maketitle

\emph{Introduction.} Spectacular advances in the quality factor of nano- and micro-mechanical oscillators and their rapidly increasing coupling to optical and microwave resonators have given rise to remarkable progress in the field of cavity optomechanics \cite{Aspelmeyer2013,Meystre2013AnnderPhysik}. This has enabled cooling of mechanical oscillators to their motional quantum ground-state \cite{Teufel2011, Chan2011} and observations of optomechanically induced transparency \cite{Weis2010Science,Teufel2011Nature_2,Safavi-Naeini2011Nature,Karuza2013PRA}, quantum zero-point motion \cite{Safavi2011, Brahms2012}, as well as squeezed light and radiation pressure shot noise \cite{Brooks2012,Purdy2013Science,Safavi-Naeini2013}. 

The interaction between light and mechanical motion due to radiation pressure is intrinsically nonlinear. While several theoretical studies of the single-photon strong-coupling regime have been reported recently \cite{Rabl2011, Nunnenkamp2011,Nunnenkamp2012,Qian2012PRL,Ludwig2012PRL, Stannigel2012PRL,Kronwald2013PRA,Liao2012}, most realizations of cavity optomechanics are still in the weak coupling limit where the coupling rate is much smaller than the cavity linewidth. Experiments to date have relied on strong optical driving, which enhances the coupling at the expense of making the effective interaction linear. Realizations that show promise for entering the strong coupling regime include the use of cold atoms \cite{Brooks2012}, superconducting circuits \cite{Teufel2011Nature_2}, microtoroids \cite{Verhagen2012Nature}, or silicon-based optomechanical crystals \cite{Chan2011}. In the latter, a ratio between the coupling rate and the cavity linewidth of 0.005 has been reported \cite{Chan2012ApplPhysLett}, and improvements seem feasible \cite{Ludwig2012PRL}. Increasing the coupling strength through collective effects in arrays of mechanical oscillators has also been proposed \cite{Xuereb2012PRL}. To enter the nonlinear regime of cavity optomechanics is of great interest, since it is only then that the internal dynamics can lead to non-classical states \footnote{By non-classical states, we mean states where the Wigner function has regions of negativity.}. 

In this article, we study corrections to linearized optomechanics and identify signatures of the intrinsic nonlinear coupling that are observable even with a relatively weak optomechanical coupling. The nonlinear effects we discuss come about due to the presence of a strong optical drive. We show that if this drive is detuned by {\it twice} the mechanical frequency from the cavity resonance frequency, two-phonon processes become resonant. This gives rise to a nonlinear version of optomechanically induced transparency (OMIT). OMIT has been well studied in linearized optomechanics \cite{Agarwal2010PRA} and is analogous to electromagnetically induced transparency in atomic systems. We point out that the two-phonon induced OMIT enables a precise measurement of the effective average phonon number of the mechanical oscillator. This provides an alternative to sideband thermometry \cite{Safavi2011, Brahms2012, Jayich2012NJP, Safavi-Naeini2013NJP}. Furthermore, we show that OMIT also occurs if the drive is detuned by {\it half} the mechanical frequency due to two-photon resonances, and the cavity response to a second probe drive is then nonlinear in probe power. We expect these effects to be observable for coupling strengths that have already been realized in experiments. Their observation would verify the intrinsic nonlinearity of the optomechanical interaction and thus open up the possibility of generating non-classical states. 

To relate to previous work, we note that a two-phonon induced transparency \cite{Huang2011PRA} can also occur in optomechanical systems where the cavity frequency depends quadratically on the position of the mechanical oscillator \cite{Nunnenkamp2010PRA}. In addition, the effect of ordinary linear OMIT on higher-order optical sidebands was studied in Ref.~\cite{Xiong2012}. 

\emph{Model}.
We consider a standard optomechanical system described by the Hamiltonian $\hat{H} = \hat{H}_\mathrm{sys} + \hat{H}_\mathrm{pump}$. The system Hamiltonian is
\begin{equation}
\label{eq:Hamiltonian}
\hat{H}_\mathrm{sys} = \hbar \omega_\C \acr \ade + \hbar \omega_\M \ccr \cde + \hbar g \left(\cde + \ccr\right) \left(\acr \ade - |\bar{a}_\Pu|^2\right) ,
\end{equation}
where $\ade \ (\cde)$ is the photon (phonon) annihilation operator, $\omega_\C \ (\omega_{\M})$ the bare cavity (mechanical) resonance frequency, and $g$ the single-photon coupling rate. The mechanical position operator is $\hat{x} =  \sqrt{\hbar/(2m \omega_\M)} \hat{z}$, where $\hat{z} = \hat{c} + \hat{c}^\dagger$ and $m$ is the effective mass. The cavity mode is driven by a laser at the frequency $\omega_\Pu$. This drive will be referred to as the pump and described by $\hat{H}_\mathrm{pump} = \im \hbar ( \ex^{-\im \omega_\Pu t} \Omega_\Pu \acr - \mathrm{h.c.})$. The constant $|\bar{a}_\Pu|^2$  in Eq.~\eqref{eq:Hamiltonian} is included for convenience and simply shifts the equilibrium position of the oscillator. We choose it to ensure that $\langle \hat{x} \rangle = 0$ in the presence of the pump, such that $\hat{x}$ is the oscillator's displacement from its average position. 

The three-wave mixing term in Eq.~\eqref{eq:Hamiltonian} is the source of the phenomena we study here, as we go beyond the usual linearization around a large cavity amplitude. Let us move to a frame rotating at the pump frequency $\omega_\Pu$ and perform a displacement transformation, such that $\ade(t) \rightarrow \ex^{-\im \omega_\Pu t} \left[\bar{a}_\Pu + \ade(t)\right]$. We define $\Delta_\Pu = \omega_\Pu - \omega_\C \neq 0$ as the pump detuning from cavity resonance and choose $\bar{a}_\Pu = \im \Omega_\Pu/\Delta_\Pu$. This results in the Hamiltonian $\hat{H} = \hat{H}_0 + \hat{H}_1$, where
\begin{eqnarray}
  \label{eq:H0}
  \hat{H}_0 & = & - \hbar \Delta_\Pu \acr \ade + \hbar \omega_\M \ccr \cde + \hbar G \left(\cde + \ccr\right) \left(\ade + \acr\right),  \\
  \hat{H}_1 & = & \hbar g \left(\cde + \ccr\right) \acr \ade .
  \label{eq:H1}
\end{eqnarray}
We have introduced $G = g \bar{a}_\Pu$ and assumed, without loss of generality, that $\bar{a}_\Pu$ is real. The coupling $G$ is enhanced by the square root of the average cavity photon number compared to $g$ and provides a bilinear coupling between photons and phonons. This coupling has been well studied, and it is known to give rise to effects such as sideband cooling \cite{Wilson-Rae2007PRL, Marquardt2007PRL, Teufel2011, Chan2011} and OMIT \cite{Agarwal2010PRA, Weis2010Science, Teufel2011Nature_2, Safavi-Naeini2011Nature}.

\emph{Identifying resonant nonlinear terms.}
The bilinear Hamiltonian $\hat{H}_0$ with $\Delta_\Pu < 0$ simply describes two linearly coupled harmonic oscillators. By a symplectic transformation, we can express $\hat{H}_0$ in terms of new operators $\hat{A}$ and $\hat{C}$, which are annihilation operators for the normal mode excitations of the system. These excitations are in general superpositions of photonic and phononic degrees of freedom. Up to a constant, the Hamiltonian becomes
\begin{equation}
  \label{eq:H0Diag}
  \hat{H}_0  =  - \hbar \tilde{\Delta}_\Pu \hat{A}^\dagger \hat{A} + \hbar \tilde{\omega}_\M \hat{C}^\dagger \hat{C} \ .
\end{equation}
We will assume that $G/\omega_\M \ll 1$, and that the pump frequency $\omega_\Pu$ does {\it not} coincide with the sideband frequencies $\omega_\C \pm \omega_\M$, but rather that $|\omega_\M \pm \Delta_\Pu|$ is on the order of $\omega_\M$. In this case, the operator $\hat{A}$ describes excitations that are photon-like, while $\hat{C}$ describes phonon-like excitations. To second order in $G/\omega_\M$, we get $\hat{A} =  [1 + 2 \lambda_+ \lambda_- \rho /(1 - \rho^2)]\ade - \lambda_+ \cde - \lambda_- \ccr - \lambda_+ \lambda_- \rho \, \acr$ and $\hat{C} = [1 + 2 \lambda_+ \lambda_- \rho /(1 - \rho^2)] \cde + \lambda_+ \ade - \lambda_- \acr + \lambda_+ \lambda_- \rho^{-1} \,  \ccr$ when we define $\rho = \omega_\M/\Delta_\Pu$ and $\lambda_\pm = G/(\Delta_\Pu \pm \omega_\M)$. The normal-mode frequencies are $\tilde{\Delta}_\Pu = \Delta_\Pu \left(1 - 2 \lambda_+ \lambda_- \rho \right)$ and $\tilde{\omega}_\M = \omega_\M \left(1 + 2 \lambda_+ \lambda_- \rho^{-1} \right)$.

We can now rewrite the Hamiltonian $\hat{H}_1$ in terms of the normal-mode operators $\hat{A}$ and $\hat{C}$, which results in multiple terms. However, since $G/\omega_\M \ll 1$, we only retain terms of nonzero order in $G/\omega_\M$ if they are resonant. We consider two different pump detunings. First, if $\Delta_\Pu \sim -2 \omega_\M$, we find $\hat{H}_1 = \hbar \, g \left(\hat{C} + \hat{C}^\dagger \right) \hat{A}^\dagger \hat{A} + \hat{H}_{1,\mathrm{res}}$, where the resonant terms are
\begin{equation}
  \label{eq:H1_2omegaM}
  \hat{H}_{1,\mathrm{res}} = \hbar  g_1 \left(\hat{A}^\dagger \hat{C}^2 + \hat{C}^{\dagger 2} \hat{A} \right) 
\end{equation}
with $g_1 = -gG/\omega_\M$. This describes processes where one photon-like excitation is created and two phonon-like excitations are destroyed, and vice versa. On the other hand, if $\Delta_\Pu \sim -\omega_\M/2$, the resonant terms are
\begin{equation}
  \label{eq:H1_omegaM2}
  \hat{H}_{1,\mathrm{res}} =  \hbar g_2 \left(\hat{A}^{\dagger  2} \hat{C} + \hat{C}^\dagger \hat{A}^2 \right)
\end{equation}
with $g_2 = - 8 g (G/\omega_\M)^2/3$, which describes processes where two photon-like excitations are created and one phonon-like excitation is destroyed, and vice versa. The Hamiltonian \eqref{eq:H0Diag} combined with \eqref{eq:H1_2omegaM} or \eqref{eq:H1_omegaM2} gives rise to new effects beyond standard linearized optomechanics. These models can be studied for a general coupling rate $g$, but we focus here on the presently experimentally relevant regime $g/\kappa \ll 1$. Specifically, we will investigate how the nonlinearities affect the response of the optical cavity to a second probe drive.

{\it Equations of motion.} We now return to the representation in terms of the original photon and phonon operators $\hat{a}$ and $\hat{c}$, and include dissipation by input-output theory \cite{Collett1984PRA, Clerk2010RMP}. The cavity and mechanical energy decay rates are $\kappa$ and $\gamma$, respectively. We assume that $\kappa \gg \gamma$ and that the system is in the resolved sideband regime where $\omega_\M > \kappa$, relevant to most experimental realizations. Note that in the presence of dissipation, the amplitude $\bar{a}_\Pu = \Omega_\Pu \chi_\C(\Delta_\Pu)$, where the cavity susceptibility is defined as $\chi_\C(\omega) = (\kappa/2 - \im \omega)^{-1}$. The drive strength $\Omega_\Pu$ is related to the laser power $P_\Pu$ through $|\Omega_\Pu|^2 = \kappa_\mathrm{ext} P_\Pu/(\hbar \omega_\Pu)$, where $\kappa_\mathrm{ext} \leq \kappa$ is the decay rate of the cavity mirror through which the cavity couples to the drive. We let $\kappa_\mathrm{int}$ describe other cavity losses, such as decay through the other mirror, scattering out of the cavity mode, absorption, etc. The sum of all decay rates equals the total cavity linewidth $\kappa = \kappa_\mathrm{ext} + \kappa_\mathrm{int}$. 

\begin{figure}
 \centering
 \includegraphics[width=0.99\columnwidth]{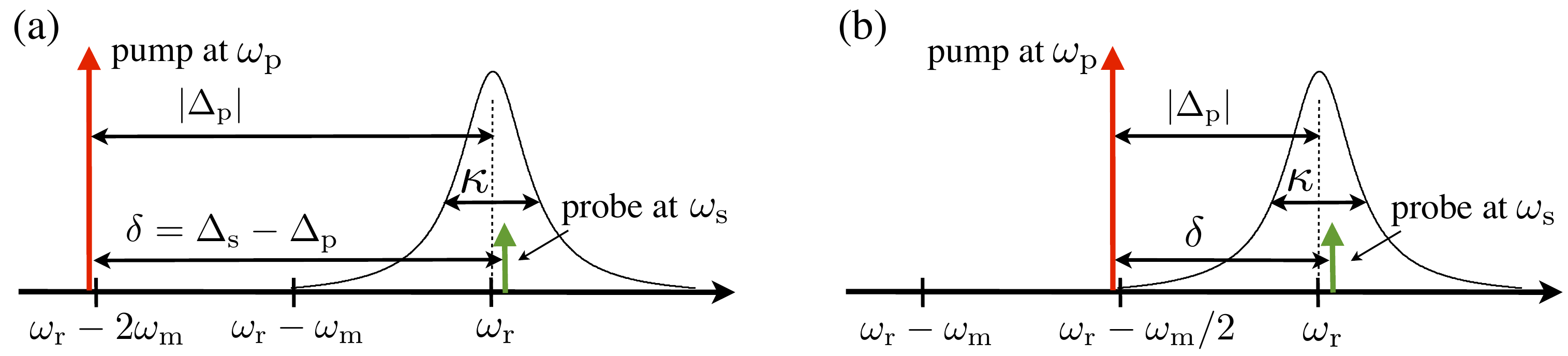}
\caption{(color online). Setup when the pump detuning is (a) $\Delta_\Pu \approx -2 \omega_\M$ and (b) $\Delta_\Pu \approx - \omega_\M/2$. Note the difference in scale between (a) and (b).}
\label{fig:setup}
\end{figure}

The quantum Langevin equations are \footnote{Since the mode hybridization is weak and the frequencies involved are only slightly renormalized, we can include dissipation in the standard way.} 
\begin{eqnarray}
  \label{eq:Langevin}
  \dot{\hat{a}} & = & - \left(\frac{\kappa}{2} - \im \Delta_\Pu\right) \hat{a} - \im (G + g \hat{a}) (\cde + \ccr) + \sqrt{\kappa} \, \hat{a}_{\mathrm{in}} \quad \ \\
  \dot{\hat{c}} & = & - \left(\frac{\gamma}{2} + \im \omega_\M\right) \cde - \im G (\ade + \acr) -\im g \hat{a}^\dagger \hat{a} + \sqrt{\gamma} \, \cde_\mathrm{in}. \quad \label{eq:Langevin2}
\end{eqnarray}
We now introduce a weak second optical drive, the probe, with frequency $\omega_\Sp$ close to the cavity resonance frequency $\omega_\C$. This is described by $\hat{H}_\mathrm{probe}(t) = \im \hbar ( \ex^{-\im \delta t} \Omega_\Sp \acr - \mathrm{h.c.})$ in the frame rotating at the pump frequency, with $\delta = \omega_\Sp - \omega_\Pu$ being the frequency difference between the probe and the pump. See Fig.~\ref{fig:setup} for an overview of the frequencies involved. 
The frequency $|\Omega_\Sp|$ is related to the probe power $P_\Sp$ by $|\Omega_\Sp|^2 = \kappa_\mathrm{ext} P_\Sp/(\hbar \omega_\Sp)$. The optical input operator in Eq.~\eqref{eq:Langevin} becomes $\sqrt{\kappa} \, \hat{a}_\mathrm{in}(t) = \ex^{-\im \delta t} \Omega_\Sp + \sqrt{\kappa_\mathrm{ext}} \xihat_\mathrm{ext}(t) + \sqrt{\kappa_\mathrm{int}} \xihat_\mathrm{int}(t)$, where the vacuum noise operators $\xihat_\mathrm{ext}$ obey $\langle \xihat_\mathrm{ext}(t) \xihat^\dagger_\mathrm{ext} (t')\rangle = \delta(t-t')$ and $\langle \xihat^\dagger_\mathrm{ext}(t) \xihat_\mathrm{ext} (t')\rangle = 0$ and similarly for $\xihat_\mathrm{int}$. The mechanical oscillator is not driven, but coupled to a thermal bath, such that the mechanical input operator obey $\langle \hat{c}_\mathrm{in}(t) \hat{c}^\dagger_\mathrm{in}(t') \rangle = (n_\therm + 1) \delta(t-t')$ and $\langle \hat{c}^\dagger_\mathrm{in}(t) \hat{c}_\mathrm{in}(t') \rangle = n_\therm \delta(t-t')$, where $n_\therm = (\ex^{\hbar \omega_\M/k_\mathrm{B} T} - 1)^{-1}$ and $T$ is the bath temperature. We will solve Eqs.~\eqref{eq:Langevin} and \eqref{eq:Langevin2} perturbatively in the single-photon coupling $g$ \footnote{The unperturbed system is stable when $G^2 < ((\kappa/2)^2 + \Delta_\Pu^2) \omega_\M/(4 |\Delta_\Pu|)$, assuming $\Delta_\Pu \leq 0$ and $\omega_\M/\gamma \gg 1$ \cite{DeJesus1987PRA}, which is satisfied here.}. The coupling $G$ cannot be treated perturbatively, but we will exploit the fact that $G/\omega_\M \ll 1$. 

The presence of two optical drives gives rise to a beat note in the optical intensity at frequency $\delta \neq \omega_\M$, and thus an off-resonant drive on the mechanical oscillator. To avoid parametric instability, the cavity frequency modulations due to the coherent motion induced by this beat note should be much smaller than the cavity linewidth, giving $g G  |\Omega_\Sp|/(\kappa \omega_\M) \ll \kappa$ by an order of magnitude estimate. This is easily fulfilled for a weak probe drive ($|\Omega_\Sp|/\kappa \sim 1$) when $G/\omega_\M, g/\kappa \ll 1$. Note that other instabilities can also arise \cite{Suh2012NanoLett} and must be avoided.

It is again convenient to move to the normal mode basis and derive Langevin equations for the operators $\hat{A}$ and $\hat{C}$. This still gives equations with linear coupling terms whenever dissipation is present. However, let us consider the extreme resolved sideband limit $\kappa/\omega_\M \ll 1$ first, where they simplify to
\begin{eqnarray}
  \label{eq:LangevinSimplify}
    \dot{\hat{A}} & = & - \left(\frac{\kappa}{2} - \im \tilde{\Delta}_\Pu\right) \hat{A} + \frac{\im}{\hbar}[\hat{H}_1,\hat{A}] + \sqrt{\kappa} \, \hat{a}_{\mathrm{in}}  \\
  \dot{\hat{C}} & = & - \left(\frac{\tilde{\gamma}}{2} + \im \tilde{\omega}_\M\right) \hat{C} + \frac{\im}{\hbar}[\hat{H}_1,\hat{C}] + \sqrt{\tilde{\gamma}} \, \tilde{c}_\mathrm{in} . \label{eq:Langevin2Simplify}
\end{eqnarray}
The effective mechanical linewidth is $\tilde{\gamma} = \gamma - \nu \kappa$ where $\nu \equiv 4 \lambda_+ \lambda_- \rho /(1 - \rho^2) < 0$ for $\Delta_\Pu < 0$. The effective frequencies $\tilde{\omega}_\M$ and $\tilde{\Delta}_\Pu$ were defined above. Note that $|\nu| \sim (G/\omega_\M)^2 \ll 1$ such that the effective mechanical linewidth is still small compared to the cavity linewidth, i.e.~$\tilde{\gamma} \ll \kappa$. The effective mechanical noise operator is defined by $\sqrt{\tilde{\gamma}} \, \tilde{c}_\mathrm{in} = \sqrt{\gamma} \, \hat{c}_\mathrm{in} + \sqrt{\kappa} (\lambda_+ \xihat + \lambda_- \xihat^\dagger)$ when ignoring the beat note and defining $\sqrt{\kappa} \, \xihat \equiv \sqrt{\kappa_\mathrm{ext}} \xihat_\mathrm{ext} + \sqrt{\kappa_\mathrm{int}} \xihat_\mathrm{int}$. Its autocorrelation properties are the same as for $\hat{c}_\mathrm{in}$, but with $n_\therm$ replaced by the effective phonon number $n_\M = (\gamma n_\therm + \kappa \lambda_-^2)/\tilde{\gamma}$. 

\emph{Two-phonon induced transparency.} We start by focusing on the case of a pump detuned by twice the mechanical frequency, $\tilde{\Delta}_\Pu = - 2 \tilde{\omega}_\M$, where two-phonon processes are resonant according to Eq.~\eqref{eq:H1_2omegaM}. Such processes have been studied before for systems with so-called quadratic optomechanical coupling \cite{Nunnenkamp2010PRA}, and it has been shown that they can lead to OMIT \cite{Huang2011PRA} much in the same way as single-phonon processes do with ordinary linear optomechanical coupling \cite{Agarwal2010PRA}. We will now see that two-phonon induced transparency can also occur in the case of linear optomechanical coupling, without the need for a nonzero quadratic coupling \footnote{Note that for a general position-dependent cavity resonance frequency $\omega_\C(x)$, the two-phonon effect we describe will dominate over that due to quadratic optomechanical coupling as long as $(\partial \omega_\C/\partial x)^2 \gg \omega_\M \partial^2 \omega_\C/\partial x^2$, a condition that is typically valid. It is usually very hard to achieve a sizable quadratic coupling, and much easier to ensure that it is small}.

By solving Eqs.~\eqref{eq:LangevinSimplify} and \eqref{eq:Langevin2Simplify} perturbatively in the single-photon coupling $g$ and transforming back to the original operators, we calculate the optical coherence $\langle \hat{a}(t) \rangle$ at frequencies close to the resonance frequency. Defining the probe beam detuning by $\Delta_\Sp = \omega_\Sp - \omega_\C$ and the effective detuning $\tilde{\Delta}_\Sp = \Delta_\Sp - \Delta_\Pu + \tilde{\Delta}_\Pu$, we find $\langle \hat{a}(t) \rangle = \ex^{-\im \delta t} \bar{a}_\Sp$ where
\begin{eqnarray}
  \label{eq:Coherence}
  \bar{a}_\Sp = \bar{a}_{\Sp,0} \left(1 - \alpha -  \frac{2 g_1^2 \, \chi_\C(\tilde{\Delta}_\Sp) \langle \hat{z}^2_0 \rangle}{\tilde{\gamma} - \im (\tilde{\Delta}_\Sp - \tilde{\Delta}_\Pu - 2 \tilde{\omega}_\M)} \right)  \ \quad
\end{eqnarray}
and $\bar{a}_{\Sp,0} = \Omega_\Sp \chi_\C(\tilde{\Delta}_\Sp)$. The first term in Eq.~\eqref{eq:Coherence} is the response of an empty cavity. The second term $\alpha$ is a small and unimportant correction due to off-resonant processes \footnote{$\alpha \approx \alpha_\mathrm{R} + \im \alpha_\mathrm{I}$ with $\alpha_\mathrm{R} = g^2 \chi_\C(\tilde{\Delta}_\Sp)[(n_\M + 1) \chi_\C(-\omega_\M) + n_\M \chi_\C(\omega_\M)]$ and $\alpha_\mathrm{I} = -2g^2 |\Omega_\Sp|^2 |\chi_\C(\tilde{\Delta}_\Sp)|^2 \chi_\C(\tilde{\Delta}_\Sp)/\omega_\M$. $\alpha_\mathrm{R}$ comes from Raman scattering of probe photons to the sideband frequencies $\omega_\mathrm{s} \pm \tilde{\omega}_\mathrm{m}$. $\alpha_\mathrm{I}$ reflects the fact that a finite number of probe photons in the cavity gives a small shift to the oscillator equilibrium position and hence the cavity resonance frequency.}. The last term gives rise to a narrow dip of width $2 \tilde{\gamma}$ in the coherent amplitude as well as a group delay of the input signal. This is analogous to the well-studied case of linear OMIT for pump detuning $\tilde{\Delta}_\Pu = - \tilde{\omega}_\M$. In the case of $\tilde{\Delta}_\Pu = - 2 \tilde{\omega}_\M$, however, the effect is {\it not} due to coherent driving of the mechanical oscillator \footnote{The linear OMIT originates from the coherent driving of the oscillator by the optical beat note. In our case, the far off-resonant drive at $\delta \sim 2 \tilde{\omega}_\M$ only leads to the shift of the cavity resonance frequency included in $\tilde{\Delta}_\Pu$ and $\tilde{\Delta}_\Sp$.}. The size of the effect rather depends on the average mechanical {\it fluctuations} through $\langle \hat{z}_0^2 \rangle \equiv (2 n_\M + 1)$. This is connected with the fact that the interaction \eqref{eq:H1} produces optical sidebands at integer multiples of $\omega_\mathrm{m}$, whose magnitudes will increase with the size of the mechanical fluctuations. Note that $\langle \hat{z}_0^2 \rangle$ can be increased by mechanically driving the oscillator. 

If the system is not in the extreme resolved sideband limit $\kappa/\omega_\M \ll 1$, Eq.~\eqref{eq:Coherence} is still valid with some corrections to the parameters, which can be found in Ref.~\cite{SM}.

\begin{figure}
 \centering
 \includegraphics[width=0.99\columnwidth,trim=2cm 1cm 3cm 0cm]{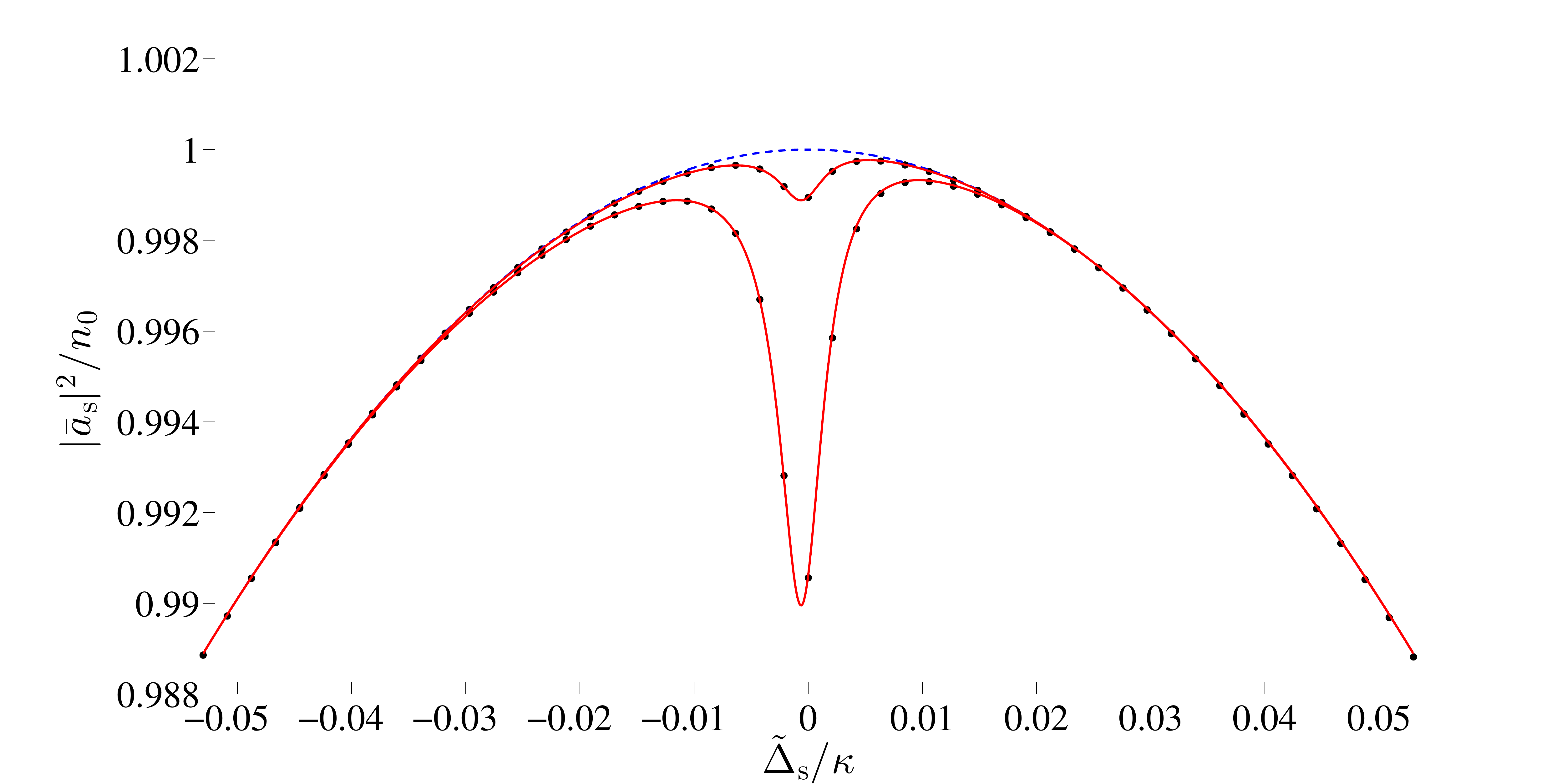}
\caption{(color online). The cavity response $|\bar{a}_\Sp|^2$ in Eq.~\eqref{eq:Coherence} in units of $n_0 = (2|\Omega_\Sp|/\kappa)^2$ for pump detuning $\tilde{\Delta}_\Pu = -2 \tilde{\omega}_\M$. The parameters are $G/\omega_\M = 0.05$, $\kappa/\omega_\M = 0.1$, $|\Omega_\Sp|/\kappa = 0.01$, $n_\therm = 1$, and $\omega_\M/\gamma = 10^5$. {\it Upper solid:} $g/\kappa = 0.01$. {\it Lower solid:} $g/\kappa = 0.03$. {\it Dashed:} $g/\kappa = 0$. {\it Dots:} Numerical results.}
\label{fig:2omegaM}
\end{figure}

The cavity response $|\bar{a}_\Sp|^2$ to the probe drive is plotted in Fig.~\ref{fig:2omegaM} for $g/\kappa = 0.01$ and $0.03$. The dip in $|\bar{a}_\Sp|^2$ corresponds to a dip in either transmission or reflection of the probe depending on the experimental setup. The parameters we used are expected to soon be within reach for silicon-based optomechanical crystals \cite{Chan2012ApplPhysLett}. We note that experimental studies of linear OMIT \cite{Weis2010Science,Teufel2011Nature_2,Safavi-Naeini2011Nature} have showed the ability to resolve dips at the percent level. Coherent interference dips are in general much easier to resolve than the incoherent noise peaks usually measured in sideband thermometry \cite{Teufel2011, Safavi2011, Brahms2012, Jayich2012NJP, Safavi-Naeini2013NJP}.

The result \eqref{eq:Coherence} provides a new way of measuring the average phonon number of the mechanical oscillator. To see this in an easy way, let us assume $\kappa/\omega_\M \ll 1$ and $\tilde{\Delta}_\Pu = -2 \tilde{\omega}_\M$, and that the mechanical oscillator is not driven. We define the dimensionless size of the dip $d \leq 1$ at $\tilde{\Delta}_\Sp = 0$ as $d \equiv 1 - |\bar{a}_\Sp/\bar{a}_{\Sp,0}(1 - \alpha)|^2  = 2 K_1 \left(2 n_\M  + 1\right)$ to lowest order in $g$, where $K_1 = 4 g_1^2/(\kappa \tilde{\gamma})$ is the {\it effective} single-photon cooperativity. In the limit where the optical broadening of the mechanical linewidth is significant, i.e.~$\kappa (G/\omega_\M)^2 \gg \gamma$, the size of the dip becomes $d = 9 (g/\kappa)^2 (2 n_\M + 1)$. We observe that the dip size $d$ {\it increases} with temperature, and does not depend on the probe drive strength $|\Omega_\Sp|$. Note that Fig.~\ref{fig:2omegaM} is the response in the low-temperature regime $n_\M \ll 1$, showing that the effect could be a useful tool for verifying ground state cooling.

The linear dependence on the oscillator fluctuations $\langle \hat{z}_0^2 \rangle$ is a result of using perturbation theory, and is only valid when $g \sqrt{\langle \hat{z}_0^2 \rangle}/\kappa \ll 1$. To gain further insight, let us consider the high-temperature regime, $n_\M \gg 1$. For $\tilde{\Delta}_\Sp = 0$ and $\tilde{\Delta}_\Pu = -2 \tilde{\omega}_\M$, a semiclassical approximation gives $\bar{a}_\Sp \approx \bar{a}_{\Sp,0}/(1 + \alpha +  K_1 \langle \hat{z}_0^2 \rangle)$, from which Eq.~\eqref{eq:Coherence} follows by expansion in $g \sqrt{\langle \hat{z}_0^2 \rangle}/\kappa$. Thus, while a dip at the percent level as in Fig.~\ref{fig:2omegaM} can be observable, the effect should be easily detectable in the high-temperature regime. For example, for an oscillator at room temperature with $\omega_\M = 2\pi \times 3$ GHz, $g/\kappa = 0.01$, $\omega_\M/\gamma = 10^5$, $\kappa/\omega_\M = 0.1$, and $G/\omega_\M = 0.05$, we get $n_\therm = 2 \times 10^3$ and $n_\M = 90$, and the dip size becomes $d = 0.14$. 

Finally, we note that while the two-phonon OMIT is a classical effect, its presence in the low-temperature limit $n_\M \rightarrow 0$ is solely due to mechanical quantum zero-point fluctuations.

\begin{figure}
 \centering
\includegraphics[width=0.99\columnwidth,trim=2cm 1cm 3cm 0cm]{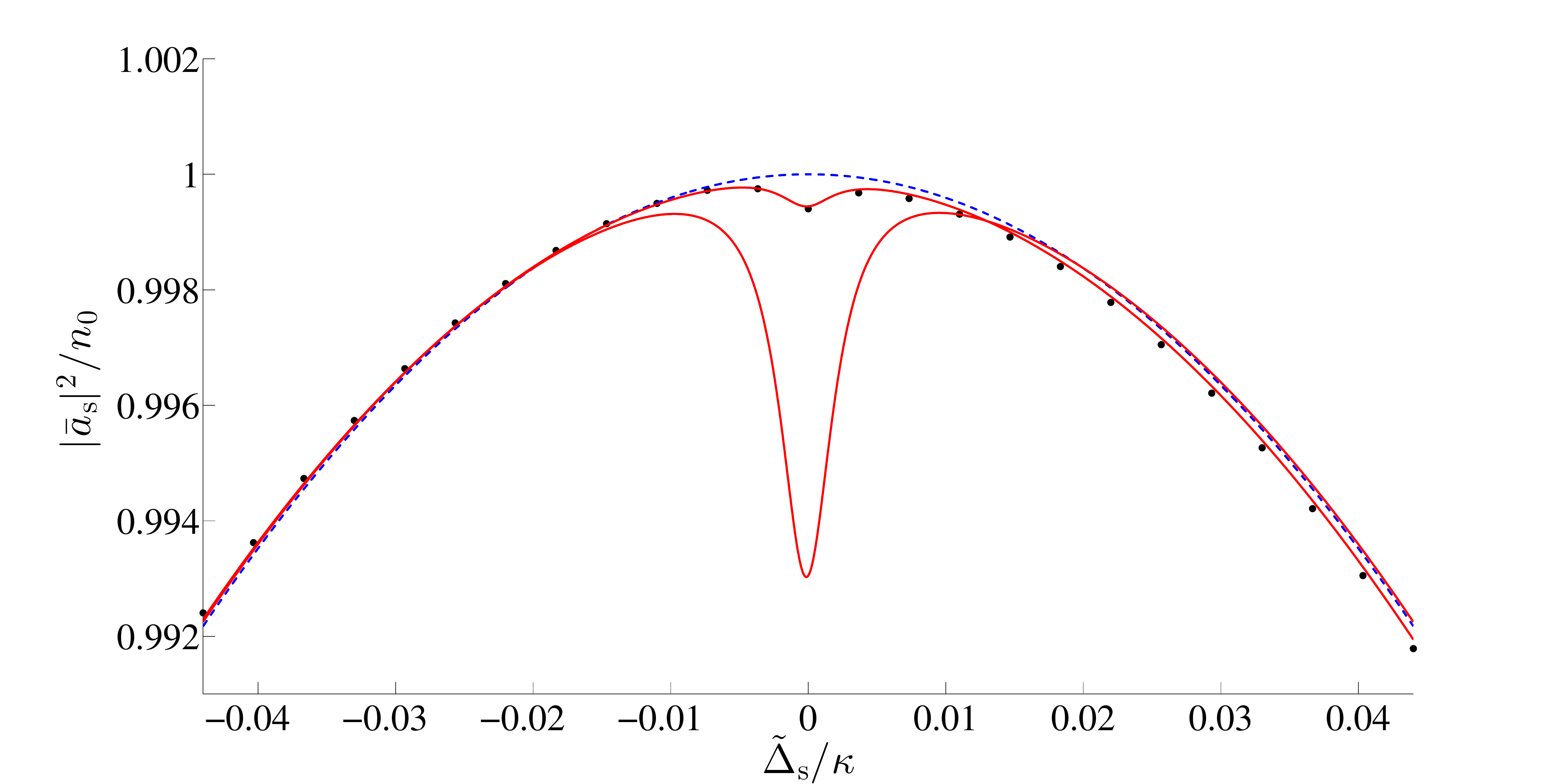}
\caption{(color online). The cavity response $|\bar{a}_\Sp|^2$ in Eq.~\eqref{eq:Coherence2} in units of $n_0 = (2|\Omega_\Sp|/\kappa)^2$ for pump detuning $\tilde{\Delta}_\Pu = -\tilde{\omega}_\M/2$. The parameters are $G/\omega_\M = 0.05$, $\kappa/\omega_\M = 0.05$, $n_\therm = 0$, and $\omega_\M/\gamma = 10^5$. {\it Upper solid:} $g/\kappa = 0.1$ and $|\Omega_\Sp|/\kappa = 0.4$.  {\it Lower solid:} $g/\kappa = 0.01$ and $|\Omega_\Sp|/\kappa = 15$. {\it Dashed:} $g/\kappa = 0$. {\it Dots:} Numerical results (only available for weak probe drives). (The small difference for larger $\tilde{\Delta}_\Sp$ comes from a difference in $\mathrm{Re} \ \alpha$.)}
\label{fig:omegaM2}
\end{figure}

{\it Two-photon induced transparency.} We now consider the case of the pump drive detuned by half the mechanical frequency, $\tilde{\Delta}_\Pu = - \tilde{\omega}_\M/2$, giving rise to the Hamiltonian \eqref{eq:H1_omegaM2}. Again, we calculate the optical coherence for frequencies close to the cavity resonance frequency, restricting ourselves to the regime $\kappa/\omega_\M \ll 1$ for simplicity. We find $\langle \hat{a}(t) \rangle   = \ex^{-\im \delta t} \bar{a}_\Sp$ with
\begin{eqnarray}
  \label{eq:Coherence2}
\bar{a}_\Sp & = & \bar{a}_{\Sp,0} \left(1 - \alpha -  \frac{2 g_2^2 \, |\bar{a}_{\Sp,0}|^2  \chi_\C(\tilde{\Delta}_\Sp)}{\tilde{\gamma}/2 - 2 \im (\tilde{\Delta}_\Sp - \tilde{\Delta}_\Pu - \tilde{\omega}_\M/2) } \right) , \quad 
\end{eqnarray}
when ignoring a very small term of order $\alpha (G/\omega_\M)^4$. There is also an OMIT effect in this case, as seen from the last term in Eq.~\eqref{eq:Coherence2}, since two probe photons can be converted to one phonon and vice versa. The dip size for $\tilde{\Delta}_\Pu = - \tilde{\omega}_\M/2$ at $\tilde{\Delta}_\Sp = 0$ becomes $d = 4 K_2 |2\Omega_\Sp/\kappa|^2 = 32 (g/\kappa)^2 (G/\omega_\M)^2 |2\Omega_\Sp/\kappa|^2$,
where the cooperativity is $K_2 = 4 g_2^2/(\kappa \tilde{\gamma})$ and the second equality assumes $\tilde{\gamma} \gg \gamma$. The amplitude $|\bar{a}_\Sp|^2$ for $\tilde{\Delta}_\Pu = - \tilde{\omega}_\M/2$ is plotted in Fig.~\ref{fig:omegaM2}. We see that even for $g/\kappa \ll 1$, the dip could be observable as it grows with increasing probe power. Note that this effect does not depend on mechanical fluctuations, but is a result of coherent motion of the oscillator at the mechanical resonance frequency induced by two-photon processes.

\emph{Numerics.}
To corroborate our analytical results, we have numerically solved the quantum master equation \cite{SM}. Figs.~\ref{fig:2omegaM} and \ref{fig:omegaM2} show that the numerical and analytical calculations are in good agreement. 

\emph{Conclusion.} We have studied corrections to linearized optomechanics and identified signatures of the intrinsic nonlinear coupling between light and mechanical motion. The signatures are nonlinear versions of optomechanically induced transparency, that come about due to resonant two-photon or two-phonon processes in the presence of a strong, off-resonant optical drive. These effects are observable even when the single-photon coupling rate is smaller than the cavity linewidth and are thus relevant to present day experiments \cite{Weis2010Science,Teufel2011Nature_2,Safavi-Naeini2011Nature}. 

\emph{Acknowledgements.}
We acknowledge financial support from The Danish Council for Independent Research under the Sapere Aude program (KB), the Swiss National Science Foundation through the NCCR Quantum Science and Technology (AN), the DARPA QuASAR program (JDT), and from the NSF under Grant No.~DMR-1004406 (SMG). The numerical calculations were performed with the Quantum Optics Toolbox \cite{Tan1999}.

\emph{Note added.}
During the final stages of this project, we became aware of related works by Lemonde, Didier, and Clerk \cite{Lemonde2013} and by Kronwald and Marquardt \cite{Kronwald2013}.



\newpage
\begin{center}
{\large {\bf Supplementary Material to ``Signatures of nonlinear cavity optomechanics in the weak coupling regime''}}\\
\end{center}

\section{Corrections outside the resolved sideband limit}

We now present the modifications to the optical response in Eq.~(11) when the system is not in the extreme resolved sideband limit $\kappa/\omega_\M \ll 1$. Additional terms must then be included in Eqs.~(9) and (10). This leads to the same expression as in Eq.~(11), but with the changes 
\begin{eqnarray}
\label{eq:corrections}
g_1^2 & \rightarrow & \im \omega_\M g^2 G^2 |\chi_\C(\omega_\M)|^2 \chi_\C(-\omega_\M) \notag \\
\langle \hat{z}_0^2 \rangle & \rightarrow & \langle \hat{z}_0^2 \rangle - \frac{\kappa}{2 \im \omega_\M} \notag \ .
\end{eqnarray}
The latter correction is due to optomechanical correlations induced by the radiation pressure shot noise. Technical laser noise will give additional corrections. The effective parameters describing the mechanical oscillator are also adjusted if $\kappa/\omega_\M$ is not negligible. For $\tilde{\Delta}_\Pu \sim - 2 \tilde{\omega}_\M$, we get 
\begin{eqnarray}
\label{eq:corrections2}
\nu & = & -8 G^2 \omega^2_\M |\chi_\C(\omega_\M)|^2 |\chi_\C(3\omega_\M)|^2 \notag \\
\tilde{\omega}_\M & = & \omega_\M - 4 G^2 \omega^3_\M [3 + (\kappa/2 \omega_\M)^2] |\chi_\C(\omega_\M)|^2 |\chi_\C(3\omega_\M)|^2  \notag \\
n_\M & = & \frac{\gamma n_\therm + \kappa G^2 |\chi_\C(3\omega_\M)|^2}{\tilde{\gamma}} \notag \ ,
\end{eqnarray}
where $\nu$ determines the effective mechanical linewidth $\tilde{\gamma}$, $\tilde{\omega}_\M$ is the effective mechanical frequency, and $n_\M$ is the average phonon number.

\section{Numerics}

To corroborate our analytical results, we numerically solve the quantum master equation  
\begin{equation}
\dot{\hat{\varrho}} = - i \left[ \op{H}(t), \hat{\varrho} \right] + \kappa \mathcal{D}[\des{a}]\hat{\varrho} + \gamma (1+n_\mathrm{th}) \mathcal{D}[\des{c}]\hat{\varrho} + \gamma n_\mathrm{th} \mathcal{D}[\cre{c}] \hat{\varrho} \notag
\label{fullmaster}
\end{equation}
for the density matrix $\hat{\varrho}(t)$ using the Hamiltonian $\hat{H}(t) = \hat{H}_0 + \hat{H}_1 + \hat{H}_\mathrm{probe}(t)$ with $\hat{H}_0$ and $\hat{H}_1$ from Eqs.~(2) and (3). Since the Hamiltonian only contains the single frequency $\delta = \omega_\Sp-\omega_\Pu$, we can use the continued-fraction method \cite{Risken1989} to solve for the frequency components of the density matrix $\op{\varrho}(t) = \sum_{n=-N}^{N} \op{\varrho}_n e^{-i n \delta t}$, where $N$ is an integer cut-off. From this, we calculate the part of the optical coherence $\langle \hat{a} \rangle$ rotating at the frequency of the probe drive. Figs.~2 and 3 show that the numerical and analytical calculations are in good agreement.

\end{document}